# Reproducibility of health claims in meta-analysis studies of COVID quarantine (stay-at-home) orders

S. Stanley Young[1] and Warren B. Kindzierski[2]

[1] CGStat, Raleigh, NC, USA
[2] Independent consultant, St Albert, Alberta, Canada

Correspondence: Warren B. Kindzierski, 12 Hart Place, St Albert, Alberta, T8N 5R1, Canada.
Email: wbk@shaw.ca or warrenk@ualberta.ca.

**Abstract**

The coronavirus pandemic (COVID) has been an extraordinary test of modern government scientific procedures that inform and shape policy. Many governments implemented COVID quarantine (stay-at-home) orders on the notion that this nonpharmaceutical intervention would delay and flatten the epidemic peak and largely benefit public health outcomes. The overall research capacity response to COVID since late 2019 has been massive. Given lack of research transparency, only a small fraction of published research has been judged by others to be reproducible before COVID. Independent evaluation of published meta-analysis on a common research question can be used to assess the reproducibility of a claim coming from that field of research. We used a p-value plotting statistical method to independently evaluate reproducibility of specific research claims made in four meta-analysis studies related to benefits/risks of COVID quarantine orders. Outcomes we investigated included: mortality, mental health symptoms, incidence of domestic violence, and suicidal ideation (thoughts of killing yourself). Three of the four meta-analyses that we evaluated (mortality, mental health symptoms, incidence of domestic violence) raise further questions about benefits/risks of this form of intervention. The fourth meta-analysis study (suicidal ideation) is unreliable. Given lack of research transparency and irreproducibility of published research, independent evaluation of meta-analysis studies using p-value plotting is offered as a way to strengthen or refute (falsify) claims made in COVID research.

## 1. Introduction
*1.1 Background*
Since late 2019, the coronavirus pandemic (COVID) has been an extraordinary test of modern government scientific procedures that inform and shape policy. Governments worldwide were faced with a disease whose severity was uncertain and was infecting millions. Governments were forced to act quickly given further uncertainties in the capacity of their health care systems to deal with the virus. In many cases, governments relied on public health experts for their policy, and more broadly to the established mechanisms by which scientific and medical expertise inform government policy.

On 11 March of 2020, the World Health Organization (WHO) officially declared COVID a pandemic (Lavezzo et al., 2020; Members, 2020). Many governments subsequently adopted aggressive pandemic policies. Examples of these policies, imposed as large-scale restrictions on people, included (Gostin et al. 2020; Jenson 2020, Magness 2021): quarantine (stay-at-home) orders, masking orders in community settings, nighttime curfews, closures of schools, universities and many businesses, and bans on large gatherings.

Mathematical modelling studies using simulated pandemic scenarios were used to justify durations of restrictions imposed on people, ranging from 2 weeks to months (CDC 2017, Jenson, 2020). These restrictions were intended to "flatten the epidemic curve" (Matrajt & Leung, 2020). The term – flatten the epidemic curve – was originally utilized by the US Centers of Disease Control for pandemic planning (CDC, 2007) to warrant use of targeted antiviral medications and nonpharmaceutical interventions (NPIs) to delay and flatten the epidemic peak.

A key aspect of flattening the epidemic curve in a pandemic was being able to spread health care demands resulting from a high incidence peak that could potentially overwhelm health care utilization capacity (Jenson, 2020). The restrictions implemented by governments, however, were lengthy as public health official policy targets shifted (Magness 2021). In United States, political influence dominated both the initiation and ultimate duration of these restrictions (Kosnik & Bellas, 2020).

*1.2 Research reproducibility*
The overall research capacity response to COVID since late 2019 has been massive (Kinsella et al., 2020; Chu et al., 2021; Ioannidis et al., 2022). To present an estimate of the magnitude of this response, we used the Advanced Search Builder capabilities of freely available PubMed search engine (pubmed.ncbi.nlm.nih.gov/advanced/). We used the terms covid[Title] OR sars-cov-2[Title] for the period 2020-2023 (search performed November 23, 2022). Our search returned 247,597 listings in the National Library of Medicine data base.

As reported in literature, only a small fraction of published research has been judged by others to be reproducible before COVID (Ioannidis, 2005, 2022; Ioannidis et al., 2011; Keown, 2012; Iqbal et al., 2016; Randall & Welser, 2018; Stodden et al., 2018). Landis et al. (2012) suggest that the inability to reproduce findings is due to a lack of research transparency.

Research transparency permits openness of study design, verification of results, synthesis of new findings with previous knowledge, and effective inquiry of research (Munafo et al., 2017). Causes of poor reproducibility of published research are related to aspects of lack of research transparency such as (Ware & Munafo, 2015): biased study designs, flexibility in research practices, low statistical power, and chasing statistical significance.

As indicated above, many research studies have been published in response to COVID. However, there remains concerns about reproducibility of COVID research, particularly where observational data are used to generate results (Bramstedt, 2020; Peng & Hicks, 2021). The current situation of irreproducible research may be that not much has changed during COVID (e.g., Gustot, 2020; Sumner et al., 2020; Paez, 2021).

*1.3 Meta-analysis*
Meta-analysis is a systematic procedure for statistically combining data (test statistics) from multiple studies that address a common research question (Egger et al., 2001), for example, whether an intervention (or risk factor) is causal of a health outcome. A meta-analysis examines a claim by taking a summary statistic along with a measure of its reliability from multiple individual intervention/risk factor—health outcome studies (called base papers) found in the literature. These statistics are combined to give what is supposed to be a more reliable estimate of an effect (Young & Kindzierski, 2019).

One aspect of replication—the performance of another study statistically confirming the same hypothesis or claim—is a cornerstone of science and replication of research claims is important before causal inference can be made (Moonesinghe et al., 2007). If a replication study result does not conform to a prevailing paradigm, it might not be submitted for publication. Also, if a similar flawed methodology is used in a replication study as in an original study, or if studies with negative findings are not submitted for publication whereas studies with positive findings are, then a false claim can be canonized (Nissen et al., 2016).

Meta-analysis has been placed at the top of the medical evidence-based pyramid – above case–control and cohort studies, and randomized trials (Murad et al., 2016). A key assumption of a meta-analysis is that estimates drawn from the base papers for the analysis are unbiased estimates of the effect of interest (Boos & Stefanski, 2013). Given these attributes, independent evaluation of published meta-analysis on a common research question can be used to assess the reproducibility of a claim coming from that field of research (Young & Kindzierski, 2019; Kindzierski et al., 2021; Young & Kindzierski, 2022a).

The objective of this study was to use a p-value plotting statistical method (after Schweder & Spjøtvoll, 1982) to independently evaluate specific research claims related to COVID quarantine (stay-at-home) orders in published meta-analysis studies. This was done in an attempt to illustrate the importance of reproducibility of research claims arising from this nonpharmaceutical intervention in the context of the surge of COVID papers in literature over the past few years.

## 2. Methods

We first wanted to gauge the number of reports of meta-analysis studies in literature related to some aspect of COVID. To do this we again used the Advanced Search Builder capabilities of the PubMed search engine. On November 20, 2022 we used the terms ((covid[Title]) OR (sars-cov-2[Title]) AND (2020:2023[pdat])) AND (meta-analysis[Title] AND (2020:2023[pdat])). Our search returned 3,204 listings in the National Library of Medicine data base. This included 633 listings for 2020, 1,301 listings for 2021, and 1,270 listings thus far for 2022. We find these counts astonishing in that a meta-analysis is a summary of available papers.

Given our understanding of pre-COVID research reproducibility of published literature discussed above, we speculated that there may be numerous meta-analysis studies relating to COVID that are irreproducible. We prepared and posted a research plan – Young & Kindzierski (2022b) – on the *Researchers.One* platform. This plan can be accessed and downloaded without restrictions from the platform. Our plan was to use p-value plotting to independently evaluate four selected published meta-analysis studies specifically relating to possible health outcomes of COVID quarantine (stay-at-home) orders – also referred to as 'lockdowns' or 'shelter-in-place' in literature.

*2.1 Data Sets*
As stated in our research plan (Young & Kindzierski, 2022b), we considered four meta-analysis studies in our evaluation:
- Herby et al. (2022) – mortality
- Prati & Mancini (2021) – psychological impacts (specifically, mental health symptoms)
- Piquero et al. (2021) – reported incidents of domestic violence
- Zhu et al. (2022) – suicidal ideation (thoughts of killing yourself)

Electronic copies of each meta-analysis study (and any corresponding electronic supplementary information files) were downloaded from the internet and read.

The Herby et al. (2022) meta-analysis examined the effect of COVID quarantine (stay-at-home) orders implemented in 2020 on mortality based on available empirical evidence. These orders were defined as the imposition of at least one compulsory, non-pharmaceutical intervention. Herby et al. initially identified 19,646 records that could potentially address their purpose.

After three levels of screening by Herby et al., 32 studies qualified. Of these, estimates from 22 studies could be converted to standardized measures for inclusion in their meta-analysis. For our evaluation, we could only consider results for 20 of the 22 studies (data they provided for two studies could not be converted to p-values). Their research claim was that "*lockdowns in the spring of 2020 had little to no effect on COVID-19 mortality*".

The Prati & Mancini (2021) meta-analysis examined the psychological impact of COVID quarantine (stay-at-home) orders on the general population. This included: mental health symptoms (such as anxiety and depression), positive psychological functioning (such as well-being and life-satisfaction), and feelings of loneliness and social support as ancillary outcomes.

Prati & Mancini initially identified 1,248 separate records that could potentially address their purpose. After screening, they identified and assessed 63 studies for eligibility and ultimately considered 25 studies for their meta-analysis. For our evaluation, we used all 20 results they reported on for mental health symptoms. Their research claim was that "*lockdowns do not have uniformly detrimental effects on mental health and most people are psychologically resilient to their effects*".

The Piquero et al. (2021) meta-analysis examined the effect of COVID quarantine (stay-at-home) orders on reported incidents of domestic violence. They used the following search terms to identify suitable papers with quantitative data to include in their meta-analysis… "domestic violence", "intimate partner violence", or "violence against women".

Piquero et al. initially identified 22,557 records that could potentially address their purpose. After screening, they assessed 132 studies for eligibility and ultimately considered 18 studies in their meta-analysis. For our evaluation, we used all 17 results (effect sizes) they reported on from the 18 studies. Their research claim was that "*incidents of domestic violence increased in response to stay-at-home/lockdown orders*".

The Zhu et al. (2021) meta-analysis examined the effect of COVID quarantine (stay-at-home) orders on suicidal ideation and suicide attempts among psychiatric patients in any setting (e.g., home, institution, etc.). They used the following search terms to identify suitable papers with quantitative data to include in their meta-analysis… "suicide" or "suicide attempt" or "suicidal ideation" or "self-harm", "psychiatric patients" or "psychiatric illness" or "mental disorders" or "psychiatric hospitalization" or "psychiatric department" or "depressive symptoms" or "obsessive-compulsive disorder".

Zhu et al. initially identified 728 records that could potentially address their purpose. After screening, they assessed 83 studies for eligibility and ultimately considered 21 studies in their meta-analysis. For our evaluation, we used all 12 results they reported on for suicidal ideation among psychiatric patients. Their research claim was that "*estimated prevalence of suicidal ideation within 12 months* [during COVID] *was… significantly higher than a world Mental Health Survey conducted by the World Health Organization (WHO) in 21 countries* [conducted 2001−2007]".

*2.2 P-value Plots*

In epidemiology it is traditional to use risk ratios and confidence intervals instead of p-values from a hypothesis test to demonstrate or interpret statistical significance. Altman & Bland (2011a,b) show that both confidence intervals and p-values are constructed from the same data and they are inter-changeable, and one can be calculated from the other.

Using JMP statistical software (SAS Institute, Cary, NC), we estimated p-values from risk ratios and confidence intervals for all data in each of the meta-analysis studies. In the case of the Herby et al. (2022) meta-analysis, standard error (SE) was presented instead of confidence intervals. Where SE values were not reports, we used the median SE of the other base studies used in the meta-analysis (6.8). The p-values for each meta-analysis are summarized in an Excel file (.xlsx format) that can be downloaded at our posted *Researchers.One* research plan (Young & Kindzierski, 2022b).

We then developed p-value plots after Schweder & Spjøtvoll (1982) to inspect the distribution of the set of p-values for each meta-analysis study. The p-value is a random variable derived from a distribution of the test statistic used to analyze data and to test a null hypothesis (Young & Kindzierski, 2022a).

In a well-designed and conducted study, the p-value is distributed uniformly over the interval 0 to 1 regardless of sample size under the null hypothesis (Schweder & Spjøtvoll, 1982). A distribution of true null hypothesis points plotted against their ranks in a p-value plot should form a 45-degree line when there are no effects (Schweder & Spjøtvoll, 1982; Hung et al., 1997; Bordewijk et al., 2020). Researchers can use a p-value plot to assess the heterogeneity of the test statistics combined in meta-analyses.

The p-value plots we constructed were interpreted as follows (Young & Kindzierski, 2022a):
- Computed p-values were ordered from smallest to largest and plotted against the integers, 1, 2, 3,…
- If p-value points on the plot followed an approximate 45-degree line, we concluded that test statistics resulted from a random (chance) process and the data supported the null hypothesis of no significant association or effect.
- If p-value points on the plot followed approximately a line with a flat/shallow slope, where most (the majority) of p-values were small ($< 0.05$), then test statistic data set provided evidence for a real, statistically significant, association or effect.
- If p-value points on the plot exhibited a bilinear shape (divided into two lines), the data set of test statistics used for meta-analysis is consistent with a two-component mixture and a general (overall) claim is not supported. In addition, a small p-value reported for the overall claim in the meta-analysis may not be valid (Schweder & Spjøtvoll, 1982).

Examples of p-value plots are provided in Appendix A after Young et al. (2022) to assist in interpretation of the p-value plots we constructed here. Specifically, the p-value plots in Appendix A represent 'plausible true null' and 'plausible true alternative' hypothesis outcomes based on published meta-analysis studies of observational data sets in the field of environmental epidemiology. As shown in the p-value plots in Appendix A:
- A plausible true null hypothesis plots as an approximate 45-degree line.
- A plausible true alternative hypothesis plots as a line with a flat/shallow slope, where most (the majority) of p-values are small (< 0.05).

The distribution of the p-value under the alternative hypothesis – where p-values are a measure of evidence against the null hypothesis – is a function of both sample size and the true value or range of true values of the tested parameter (Hung et al., 1997). The p-value plots presented in Young et al. (2022) represent examples of distinct (single) sample distributions for each condition – i.e., for true null associations and true effects between two variables. Evidence for p-value plots exhibiting behaviors outside of that shown in Young et al. (2022) should initially be treated as ambiguous (uncertain).

## 3. Results

Mortality
Our independent evaluation of the effect of COVID quarantine (stay-at-home) orders on mortality – the Herby et al. (2022) meta-analysis – is shown in Figure 1. There are 20 studies that we included in the figure. Six of the 20 studies had p-values below 0.05 while four of the studies had p-values close to 1.00. Ten studies fell roughly on a 45-degree line implying random results.

This data set comprises mostly null associations (14) and with five or six possible associations with effects (1-in-20 could be chance, false, positive association). While not ideal, this data set is a closer fit to a sample distribution for a true null association between two variables. Our interpretation of the p-value plot is that COVID quarantine (stay-at-home) orders are not supported for reducing mortality, consistent with Herby et al. (2022).

[Fig 1 to be inserted here]
*Figure 1. P-value plot (p-value versus rank) for Herby et al. (2022) meta-analysis of the effect of COVID quarantine (stay-at-home) orders implemented in 2020 on mortality. Symbols (circles) are p-values ordered from smallest to largest (n=20).*

Psychological impact (mental health symptoms)
Our independent evaluation of the effect of COVID quarantine (stay-at-home) orders on mental health symptoms – the Prati & Mancini (2021) meta-analysis – is shown in Figure 2. Figure 2 presents as a bilinear shape showing a two-component mixture. This data set clearly does not represent a distinct sample distribution for either true null associations or true effects between two variables. Our interpretation of the p-value plot is that COVID quarantine (stay-at-home) orders have an ambiguous (uncertain) effect on mental health symptoms. However as discussed below, there are valid questions their research claim.

[Fig 2 to be inserted here]
*Figure 2. P-value plot (p-value versus rank) for Prati & Mancini (2021) meta-analysis of the effect of COVID quarantine (stay-at-home) orders on mental health symptoms. Symbols (circles) are p-values ordered from smallest to largest (n=20).*

Incidents of domestic violence
Our independent evaluation of the effect of COVID quarantine (stay-at-home) orders on reported incidents of domestic violence – the Piquero et al. (2021) meta-analysis – is shown in Figure 3. Thirteen of the 17 studies had p-values less than 0.05. While not shown in the figure, eight of the p-values were small (<0.001).

This data set comprises mostly non-null associations (13) and with four possible null associations. While not perfect, this data set is a closer fit to a sample distribution for a true alternative association between two variables. Our interpretation of the p-value plot is that COVID quarantine (stay-at-home) have a negative effect (increase) for reported incidents of domestic violence.

[Fig 3 to be inserted here]
*Figure 3. P-value plot (p-value versus rank) for Piquero et al. (2021) meta-analysis of the effect of COVID quarantine (stay-at-home) orders on reported incidents of domestic violence. Symbols (circles) are p-values ordered from smallest to largest (n=17).*

Suicidal ideation
Our independent evaluation of the effect of COVID quarantine (stay-at-home) orders on suicidal ideation – the Zhu et al. (2021) meta-analysis – is shown in Figure 4. The p-values for all 12 studies were less than 0.05. Ten of the 12 studies had p-values less than 0.05. While not shown in the figure, eight of the p-values were small (<0.001).

This data set presents as a distinct sample distribution for true effects between two variables. Our interpretation of the p-value plot is that COVID quarantine (stay-at-home) orders have an effect on suicidal ideation (thoughts of killing yourself). However as discussed below, there are valid questions about how the meta-analysis was formulated.

[Fig 4 to be inserted here]
*Figure 4. P-value plot (p-value versus rank) for Zhu et al. (2021) meta-analysis of the effect of COVID quarantine (stay-at-home) orders on suicidal ideation (thoughts of killing yourself). Symbols (circles) are p-values ordered from smallest to largest (n=12).*

## 4. Discussion

As stated previously, independent evaluation of published meta-analysis on a common research question can be used to assess the reproducibility of a claim coming from that field of research. We evaluated four meta-analysis studies of COVID quarantine (stay-at-home) orders implemented in 2020 and corresponding health benefits and/or harms. Our intent was to illustrate the importance of reproducibility of research claims arising from this nonpharmaceutical intervention in the context of the surge of COVID papers in literature over the past few years.

Mortality
The Herby et al. (2022) meta-analysis examined the effect of COVID quarantine orders on mortality. Their research claim was that "*lockdowns in the spring of 2020 had little to no effect on COVID-19 mortality*". Here, they imply that the intervention (COVID quarantine orders) had little or no effect on reduction of mortality.

The quantitative data Herby et al. present to put their findings into perspective is that they estimated the average lockdown in United States (Europe) in the spring of 2020 avoided 16,000 (23,000) deaths. In contrast, they report that there are about 38,000 (72,000) flu deaths occurring each year in the United States (Europe).

Our evidence agrees with their claim. Our p-value plot (Figure 1) is not consistent with expected behaviour of a distinct sample distribution for a true effect between the intervention (quarantine) and the outcome (reduction in mortality). More importantly, our plot shows considerable randomness (many null associations, p-values > 0.05) supporting no consistent effect. Herby et al. further stated that "*costs to society must be compared to the benefits of lockdowns, which our meta-analysis has shown are little to none*".

Psychological impact (mental health symptoms)
The Prati & Mancini (2021) meta-analysis examined the psychological impact of COVID quarantine orders on the general population. Their research claim was that "*lockdowns do not have uniformly detrimental effects on mental health and most people are psychologically resilient to their effects*". We evaluated a component of psychological impact – i.e., whether COVID quarantine orders affect mental health symptoms (Figure 2). Figure 2 clearly exhibits a two-component mixture implying an ambiguous (uncertain) effect on mental health symptoms. However, our evidence does not necessarily support their claim.

Digging deep into their study reveals an interesting finding. Their study looked at a variety of psychological symptoms that differed from study to study. Although not shown here, when they examined these symptoms separately – a meta-analysis of each symptom – there was a strong signal for anxiety (p-value less than 0.0001). This is less than a Boos & Stefanski (2011) proposed p-value action level of 0.001 for expected replicability. Here, the term 'action level' means that if a study is replicated, the replication will give a p-value less than 0.05.

We also note that Prati & Mancini appear to take absence of evidence of a negative mental health effect of COVID quarantine orders in their meta-analysis as implying it does not affect mental health. But absence of evidence does not imply evidence of absence (Altman & Bland, 1995, Alderson, 2004; Sedgwick, 2014). Just because meta-analysis failed to find an effect, it does not imply that "…*most people are psychologically resilient to their* [lockdown] *effects*". A more plausible and valid inference is that this statement of claim is insufficiently researched at this point.

Incidents of domestic violence
The Piquero et al. (2021) meta-analysis examined COVID quarantine orders on reported incidents of domestic violence. Their research claim was that "*incidents of domestic violence increased in response to stay-at-home/lockdown orders*". Our evidence suggests agreement with this claim. Our p-value plot (Figure 3) is more consistent with expected behaviour of a distinct sample distribution for a true effect between the intervention (quarantine) and the outcome (increase in incidents of domestic violence).

Several null association studies exist within their data set. We note that Figure 3 has 13 of 17 p-values less than 0.05, with eight of these less than 0.001. Our evidence supports that COVID quarantine orders <u>likely</u> increased incidents of domestic violence.

Suicidal ideation
The Zhu et al. (2021) meta-analysis examined COVID quarantine orders on suicidal ideation (thoughts of killing yourself). Their research claim was that "*estimated prevalence of suicidal ideation within 12 months* [during COVID] *was… significantly higher than a world Mental Health Survey conducted by the World Health Organization (WHO) in 21 countries* [conducted 2001−2007]".

The p-value plot (Figure 4) strongly supports their claim. The plot is very consistent with expected behaviour of a distinct sample distribution for a true effect between the intervention (quarantine) and the outcome (increased prevalence of suicidal ideation). However, digging deep into their study reveals a problem in the formulation of their meta-analysis.

In strong science, a research question being investigated is judged against a control. Zhu et al. effectively ignores controls in their meta-analysis. They compared incidence of suicidal ideation against a zero standard and not to control groups. Specifically, the pre-COVID (i.e., background) suicidal ideation signal is ignored in their meta-analysis.

Indeed, in their Table 1 they present results from the base papers where data for control groups is available. For example, the Seifert et al. (2021) base paper notes suicidal ideation presented in 123 of 374 patients in the psychiatric emergency department of Hannover Medical School during the pandemic, and 141 of 476 in the same department before the pandemic – 32.9%versus 29.6%. The difference is not significant.

Comparing their Table 1 data set with their Figure 1 forest plot, Zhu et al. only carried 32.9% into their meta-analysis, in effect ignoring the control data. It is the same situation with all data set entries in their Figure 1. Zhu et al. only considered pandemic incidence in their meta-analysis, and they ignored any control data. How they formulated their work calls their claims into serious question. We conclude that the Zhu et al. results are unreliable.

Implications
COVID quarantine orders were implemented on the notion that this nonpharmaceutical intervention would delay and flatten the epidemic peak and benefit public health outcomes overall. Three of the four meta-analyses that we evaluated raise questions about public health benefits/risks of this form of nonpharmaceutical intervention. The fourth meta-analysis study is unreliable.

One meta-analysis that we evaluated – Herby et al. (2022) – questions the benefits of this form of intervention for preventing mortality. Our p-value plot supports their finding that COVID quarantine orders had little or no effect on reduction of mortality.

A second meta-analysis – Prati & Mancini (2021) assessment of mental health symptoms – offers confounding evidence. Our p-value plot clearly exhibits a two-component mixture implying an ambiguous (uncertain) effect between COVID quarantine orders and mental health symptoms. However, data for a component of mental health symptoms (anxiety) suggests a negative effect from COVID quarantine orders. Further, Prati & Mancini (2021) lack evidence to claim that "…*most people are psychologically resilient to their* [lockdown] *effects*".

Our evaluation of the Piquero et al. (2021) meta-analysis – assessment of domestic violence incidents – supports a true effect between the intervention (quarantine) and the outcome (increase in incidents of domestic violence) with additional confirmatory research needed. Finally, the meta-analysis of Zhu et al. (2021) on suicidal ideation (thoughts of killing yourself) is wrongly formulated and should be disregarded until or unless controls are included in the analysis.

Standing back and looking at the overall findings of these studies, benefits of COVID quarantine orders remain uncertain and risks (negative public health consequences) of this intervention cannot be ruled out. Given that the base studies and the meta-analyses themselves were, for the most part, rapidly conducted and published, we acknowledge that confirmatory research for some of the outcomes investigated is warranted.

Our interpretation of COVID quarantine benefits/risks is consistent, for example, with earlier research of James (2020) and conventional wisdom, Inglesby et al. 2006. James takes a position that is it unclear whether there were benefits from this intervention relative to less restrictive measures aimed at controlling "risky" personal interactions (e.g., mass gatherings and large clusters of individuals in enclosed spaces).

James (2020) also notes numerous economic and public health harms in the United States as May 1, 2020:
- Over 20 million newly unemployed.
- State-wide school closures across the country.
- Increased spouse and child abuse reports.
- Increased divorces.
- Increased backlog of patient needs for mental health services, cancer treatments, dialysis treatments and everyday visits for routine care.
- Increased acute emergency services.

This is consistent with interim quantitative data as of September 2020 presented by the American Institute of Economic Research (2020) on the cost and negative public health implications of pandemic restrictions in United States and around the world.

**Acknowledgments**
No external funding was provided for this study. The study was conceived based on previous work undertaken by CG Stat for the National Association of Scholars (nas.org), New York, NY.

Figures

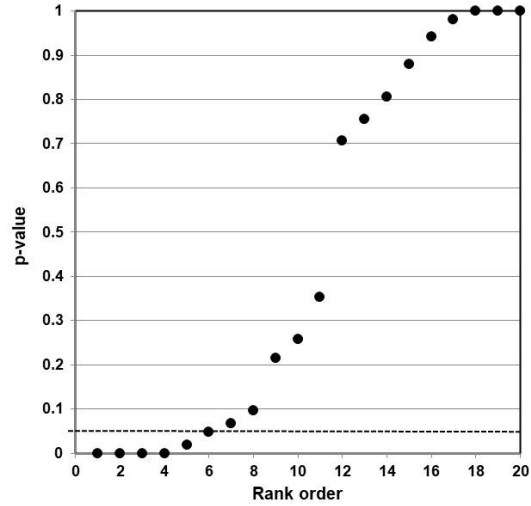

Figure 1. P-value plot (p-value versus rank) for Herby et al. (2022) meta-analysis of the effect of COVID quarantine (stay-at-home) orders implemented in 2020 on mortality. Symbols (circles) are p-values ordered from smallest to largest (n=20).

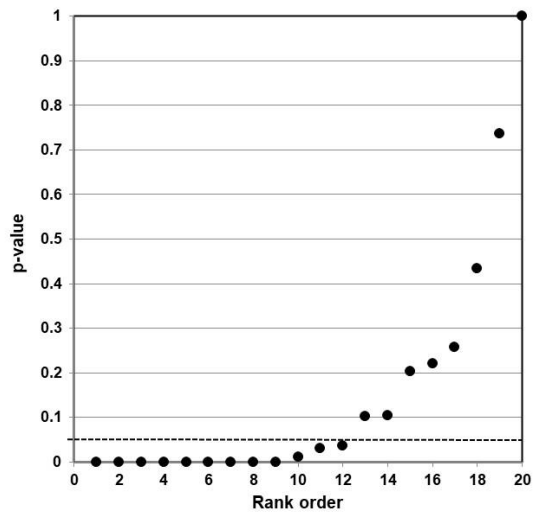

Figure 2. P-value plot (p-value versus rank) for Prati & Mancini (2021) meta-analysis of the effect of COVID quarantine (stay-at-home) orders on mental health symptoms. Symbols (circles) are p-values ordered from smallest to largest (n=20).

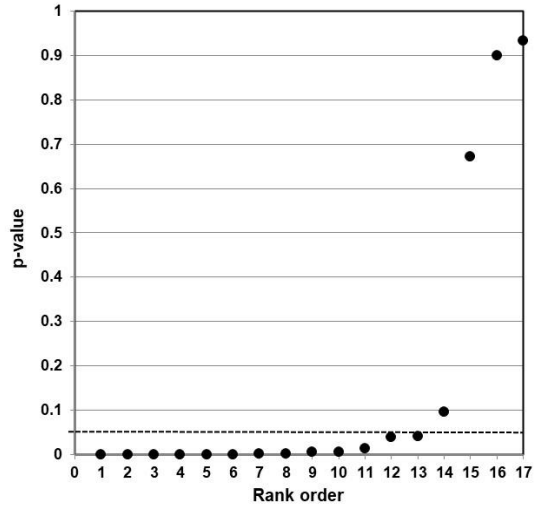

Figure 3. P-value plot (p-value versus rank) for Piquero et al. (2021) meta-analysis of the effect of COVID quarantine (stay-at-home) orders on reported incidents of domestic violence. Symbols (circles) are p-values ordered from smallest to largest (n=17).

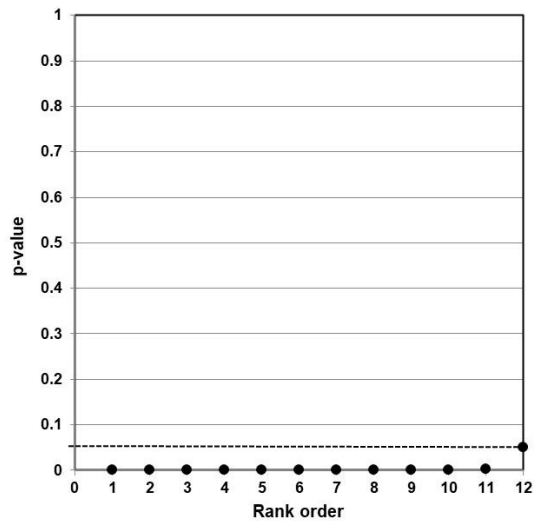

Figure 4. P-value plot (p-value versus rank) for Zhu et al. (2021) meta-analysis of the effect of COVID quarantine (stay-at-home) orders on suicidal ideation (thoughts of killing yourself). Symbols (circles) are p-values ordered from smallest to largest (n=12).